# Enhancement of the Raman Effect by Infrared Pumping

V. Yu. Shishkov[1,2,3], E. S. Andrianov[1,2], A. A. Pukhov[1,2,3], A. P. Vinogradov[1,2,3], and A. A. Lisyansky[4,5]

[1]Dukhov Research Institute of Automatics (VNIIA), 22 Sushchevskaya, Moskow 127055, Russia
[2]Moscow Institute of Physics and Technology, 9 Institutskiy per., Dolgoprudny 141700, Moscow reg., Russia
[3]Institute for Theoretical and Applied Electromagnetics, 13 Izhorskaya, Moscow 125412, Russia
[4]Department of Physics, Queens College of the City University of New York, Flushing, New York 11367, USA
[5]The Graduate Center of the City University of New York, New York, New York 10016, USA

We propose a method for increasing the Raman scattering from an ensemble of molecules by up to four orders of magnitude. Our method requires an additional coherent source of IR radiation with the half-frequency of the Stokes shift. This radiation excites the molecule electronic subsystem that in turn, via Fröhlich coupling, parametrically excites nuclear oscillations at a resonant frequency. This motion is coherent and leads to a boost of the Raman signal in comparison to the spontaneous signal because its intensity is proportional to the squared number of molecules in the illuminated volume.

**Introduction**

Since the discovery of the Raman effect in 1928 [1, 2], the Raman spectroscopy has become the main tool for studying molecular vibrations [3]. The frequency of the Raman peaks and their widths are determined by the internal structure of the molecule providing information that is used in various applications such as bioimaging [4-8], sensing [9-11], temperature measurements [12-17], and in the research of two-dimensional materials [18, 19].

The main disadvantage of the Raman spectroscopy is that spontaneous Raman signals are weak. This weakness is caused by the smallness of the indirect interaction of optical waves with vibrations of molecules. This limits the minimum concentration of molecules of the analyte and the minimum source intensity required for the spectroscopy.

There are several effective methods for enhancing Raman signals. Surface-enhanced Raman scattering (SERS) [20-22] and tip-enhanced Raman spectroscopy (TERS) [9, 23, 24] utilize the enhancement of the local field around a molecule using the plasmon resonance [25-27]. Other methods employ the parametric excitation of nucleus vibrations in the molecule. This approach is used in coherent anti-Stokes Raman scattering (CARS) [28-30], in coherent Stokes Raman scattering (CSRS) [31-34] as well as in surface-enhanced coherent anti-Stokes Raman scattering (SECARS) [35]. Obtaining an enhancement in CARS requires strong driving fields. Consequently, some undesirable side effects arise; in particular, the stimulated Raman scattering (SRS) can result in the energy transition between two laser beams that leads to the signal suppression and limits the sensitivity [36]. Therefore, a method that on the one hand, similar to CARS, allows for coherent excitation of oscillations of molecular nuclei, but on the other hand, does not suffer from a strong energy transition between waves is highly desirable.

In this paper, we propose a method for enhancing Raman signals from molecules at both Stokes and anti-Stokes frequencies. This method utilizes an additional coherent infrared (IR) source. Below we consider molecules, which transitions between vibrational states have zero dipole moments. Consequently, the IR radiation can directly affect the electronic subsystem only. The influence on the vibrational motion of nuclei is realized via the interaction between electronic and nucleus subsystems of the molecule. This interaction, which may be represented in the form of the Fröhlich interaction Hamiltonian [37], results in an effective force affecting the nucleus vibrations. The frequency spectrum of this force contains all possible sums and differences of frequencies of participating waves. In our work, a doubled IR frequency may coincide with the eigenfrequency of vibrational states causing a resonant excitation of nucleus vibrations. Due to the coherence of the IR radiation, the excited vibrational motions of nuclei in different molecules are also coherent leading to coherence in signals from the molecules at both Stokes and anti-Stokes frequencies. This results in an enhancement of the Raman signal proportional to the squared number of molecules in the pumped volume while the Raman signal originated from incoherent vibrational nucleus motion is linearly proportional to the number of molecules.

**Infrared Pumping**

For molecules with zero dipole moments of the transitions between vibrational states, light may only interact with the electronic subsystem of a molecule. The excitation of nuclear oscillations may arise due to the interaction between electron and nucleus vibrations. The nature of this interaction is the same as the interaction of the electronic subsystem with vibrational motion of nuclei at the Raman scattering. We take into account only one vibrational mode, which is considered in the harmonic approximation. Illumination of a molecule by IR light results in mixing of ground and excited states of the electronic subsystem. This causes a periodic displacement of negative charges [37], which, in turn, results in a periodic motion of molecule nuclei that tends to keep local equilibrium. The electronic subsystem of a molecule can be treated as an effective two-level system (TLS), and one can show [37] that the Hamiltonian of the interaction between the electronic subsystem and vibrations of nuclei takes the form of the Fröhlich Hamiltonian [the third term in Eq. (1)]. Such a system can be described with the Heisenberg-Langevin equations [38, 39]. To obtain these equations, we use the Hamiltonian [40-46]:

$$\hat{H}_S = \hbar\omega_0\hat{\sigma}^\dagger\hat{\sigma} + \hbar\omega_v\hat{b}^\dagger\hat{b} + \hbar g\hat{\sigma}^\dagger\hat{\sigma}\left(\hat{b}^\dagger + \hat{b}\right) + \hbar\left(\hat{\sigma}^\dagger + \hat{\sigma}\right)\Omega_{vis}\cos(\omega_{vis}t) + \hbar\left(\hat{\sigma}^\dagger + \hat{\sigma}\right)\Omega_{IR}\cos(\omega_{IR}t), \quad (1)$$

where $\omega_{vis}$, $\omega_0$, and $\omega_v$ are the frequencies of the incident optical light, the TLS transition, and vibrational motion, respectively. We consider a non-resonant case, $|\omega_0 - \omega_{vis}| \gg \gamma_{\perp 0}$, where $\gamma_\perp$ is the TLS transverse relaxation rate. $\hat{\sigma}$ and $\hat{\sigma}^\dagger$ are the transition operators of the TLS, $\omega_v$ is the vibrational frequency of nuclei, $\hat{b}$ and $\hat{b}^\dagger$ are the creation and annihilation operators of the vibrational motion, $\omega_{IR}$ is the frequency of the IR source, and $g$ is the coupling constant of electrons and vibrations of molecule nuclei. The last two terms of Eq. (1) describe the interaction between the electronic subsystem of the molecule with the external fields; the Rabi constants of these interactions are $\Omega_{vis} = -\mathbf{d}_{eg} \cdot \mathbf{E}_{vis} / \hbar$ and $\Omega_{IR} = -\mathbf{d}_{eg} \cdot \mathbf{E}_{IR} / \hbar$, respectively, where $\mathbf{d}_{eg}$ is the TLS transition dipole moment, $\mathbf{E}_{vis}$ and $\mathbf{E}_{IR}$ are the electric field amplitudes of the visible and IR incident lights.

Using Hamiltonian (1) one can obtain the Heisenberg-Langevin equations [39, 47] for the TLS and operators of the vibrational motion (see Supplemental Material for details):

$$d\hat{b}/dt = (-i\omega_v - \gamma_v)\hat{b} - ig\hat{\sigma}^\dagger\hat{\sigma} + \hat{F}_b(t), \quad (2)$$

$$d\hat{\sigma}/dt = (-i\omega_0 - \gamma_\perp)\hat{\sigma} + i(2\hat{\sigma}^\dagger\hat{\sigma} - 1)(\Omega_{vis}\cos(\omega_{vis}t) + \Omega_{IR}\cos(\omega_{IR}t)) - ig\hat{\sigma}(\hat{b}^\dagger + \hat{b}) + \hat{F}_\sigma(t), \quad (3)$$

where $\gamma_v$ is the decay rate of the amplitude of vibrations, $\hat{F}_b(t)$ and $\hat{F}_\sigma(t)$ are the noise operators for the vibrations and TLS, respectively. Dissipation and noise are not independent but are related to each other by the fluctuation-dissipation theorem [39, 48]. We consider the case of $\hbar\omega_0 \gg kT$, which is common in quantum optics. Then, we can omit the TLS noise by putting $\hat{F}_\sigma(t) = 0$. The noise $\hat{F}_b(t)$ is the thermal noise with the correlations $\langle \hat{F}_b^\dagger(t_1)\hat{F}_b(t_2)\rangle = \gamma_v \bar{n}_v \delta(t_1 - t_2)$ and $\langle \hat{F}_b(t_1)\hat{F}_b^\dagger(t_2)\rangle = \gamma_v (1 + \bar{n}_v)\delta(t_1 - t_2)$ with $\bar{n}_v = (\exp(\hbar\omega_v/kT) - 1)^{-1}$.

Let us introduce the parameter

$$\varepsilon = max\{\Omega_{vis}/|\omega_{vis} - \omega_0|, \Omega_{IR}/|\omega_{IR} - \omega_0|\}, \quad (4)$$

where $\Omega_{vis}/|\omega_{vis} - \omega_0| \ll 1$ and $\Omega_{IR}/|\omega_{IR} - \omega_0| \ll 1$. Smallness of $\varepsilon$ allows for using the perturbation method for solving Eqs. (2) and (3). Below we use the series expansion of the averages of the operators $\hat{\sigma}$ and $\hat{b}$, $\hat{\sigma} = \hat{\sigma}_0 + \hat{\sigma}_1 + ...$ and $\hat{b} = \hat{b}_0 + \hat{b}_1 + ...$.

In the zeroth order of the perturbation theory, we have

$$d\hat{b}_0/dt + (i\omega_v + \gamma_v)\hat{b}_0 = \hat{F}_b(t), \quad (5)$$

$$d\hat{\sigma}_0/dt + (i\omega_0 + \gamma_\perp)\hat{\sigma}_0 = 0, \tag{6}$$

The solutions for these equations are

$$\hat{b}_0(t) = \int_0^t G_b(t-t')\hat{F}_b(t')dt', \tag{7}$$

$$\hat{\sigma}_0 = 0, \tag{8}$$

where $G_b(t-t') = \exp((-i\omega_v - \gamma_v)(t-t'))$. Note that $\hat{b}_0(t)$ describes incoherent, thermal vibrational motion with $\langle \hat{b}_0^\dagger(t)\hat{b}_0(t)\rangle = \bar{n}_v = (\exp(\hbar\omega_v/kT) - 1)^{-1}$.

In the first order of the perturbation theory, we obtain

$$d\hat{b}_1/dt + (i\omega_v + \gamma_v)\hat{b}_1 = 0, \tag{9}$$

$$d\hat{\sigma}_1/dt + (i\omega_0 + \gamma_\perp)\hat{\sigma}_1 = -i\Omega_{vis}\cos(\omega_{vis}t) - i\Omega_{IR}\cos(\omega_{IR}t). \tag{10}$$

The solution to Eqs. (9) and (10) are

$$\hat{b}_1(t) = 0 \tag{11}$$

$$\hat{\sigma}_1(t) = \frac{1}{2}\frac{\Omega_{vis}}{\omega_{vis} - \omega_0}e^{-i\omega_{vis}t} + \frac{1}{2}\frac{\Omega_{vis}}{\omega_{vis} + \omega_0}e^{i\omega_{vis}t} + \frac{1}{2}\frac{\Omega_{IR}}{\omega_{IR} - \omega_0}e^{-i\omega_{IR}t} + \frac{1}{2}\frac{\Omega_{IR}}{\omega_{IR} + \omega_0}e^{i\omega_{IR}t}, \tag{12}$$

In obtaining Eq. (12) we take into account that the frequencies of the external fields are far from the TLS transition frequency, $|\omega_{vis} - \omega_0| \gg \gamma_\perp$ and $|\omega_{IR} - \omega_0| \gg \gamma_\perp$. We, therefore, neglect the imaginary part in the denominator of Eq. (12).

The first order of the perturbation theory, Eq. (12), describes the Rayleigh scattering but it still does not describe the Raman effect. In the second order of the perturbation theory, we obtain

$$d\hat{b}_2/dt + (i\omega_v + \gamma_v)\hat{b}_2 = -ig\hat{\sigma}_1^\dagger\hat{\sigma}_1 \tag{13}$$

$$d\hat{\sigma}_2/dt + (i\omega_0 + \gamma_\perp)\hat{\sigma}_2 = -2ig\hat{\sigma}_1(t)(\hat{b}_0^\dagger(t) + \hat{b}_0(t)). \tag{14}$$

The right-hand part of Eq. (13) can be considered as an effective force. According to Eq. (12), this force has the form of a sum of products weighted cosines, $\cos(2\omega_{vis}t)$, $\cos((\omega_{IR} - \omega_{vis})t)$, $\cos((\omega_{IR} + \omega_{vis})t)$, $\cos(2\omega_{IR}t)$, and 1 [weighting coefficients are derived in the Supplemental

Material, Eqs. (S14)-(S18)]. The resonant excitation of the vibrational motion occurs if one of the following relations are fulfilled $2\omega_{vis} = \omega_v$, $\omega_{vis} - \omega_{IR} = \omega_v$, $\omega_{IR} + \omega_{vis} = \omega_v$, and $2\omega_{IR} = \omega_v$. If $\omega_{IR}$ were in the optical range, the second relationships was equivalent to CARS, and the others could not be realized. We are interested in the case when $\omega_{IR}$ is in the IR range.

The second harmonics in the right-hand side of the Eq. (13), arising due to the nonlinearity of the Fröhlich Hamiltonian, is responsible for the resonant excitation of vibrations when the frequency of the IR radiation is $\omega_{IR} \approx \omega_v / 2$. Implying that $\omega_{IR} \approx \omega_v / 2$, in the right-hand side of Eq. (13), we retain the resonant term only. Since this term is independent of incoherent variables, in particular of $\hat{b}_0(t)$, it leads to the excitation of the coherent vibrational motion: in deriving Eqs. (13) and (14) we disregard nonresonant terms. Equation (13) takes the form

$$d\hat{b}_2/dt + (i\omega_v + \gamma_v)\hat{b}_2 \approx i\frac{g}{4}\frac{\Omega_{IR}^2}{\omega_0^2 - \omega_{IR}^2}\exp(-i2\omega_{IR}t) \tag{15}$$

The solutions of Eqs. (14)-(15) are

$$\hat{b}_2 \approx \frac{1}{4}\frac{g}{(\omega_v - 2\omega_{IR}) - i\gamma_v}\frac{\Omega_{IR}^2}{\omega_0^2 - \omega_{IR}^2}e^{-i2\omega_{IR}t}, \tag{16}$$

$$\hat{\sigma}_2(t) = -2ig\int_0^t dt'' G_\sigma(t-t'')\hat{\sigma}_1(t'')\operatorname{Re}\int_0^{t''} dt' G_b(t''-t')\hat{F}_b(t') \tag{17}$$

where $G_\sigma(t-t') = \exp((-i\omega_0 - \gamma_\perp)(t-t'))$ is the Green function for Eq. (14). The right-hand side of Eq. (17) is due to the influence of incoherent vibrational motion on the electronic subsystem; this term leads to the spontaneous Raman scattering [37, 45, 49].

The second order of the perturbation theory, Eq. (17), describes the spontaneous Raman scattering near the frequencies $\omega_{vis}$ and $\omega_{IR}$. Indeed, summing up terms for $\hat{\sigma}$ from Eqs. (8), (12), and (17) we obtain

$$\hat{\sigma}(t) = \frac{1}{2}\frac{\Omega_{vis}}{\omega_{vis} - \omega_0}e^{-i\omega_{vis}t} + \frac{1}{2}\frac{\Omega_{vis}}{\omega_{vis} + \omega_0}e^{i\omega_{vis}t} + \frac{1}{2}\frac{\Omega_{IR}}{\omega_{IR} - \omega_0}e^{-i\omega_{IR}t} + \frac{1}{2}\frac{\Omega_{IR}}{\omega_{IR} + \omega_0}e^{i\omega_{IR}t}$$
$$-2ig\int_0^t dt'' G_\sigma(t-t'')\hat{\sigma}_1(t'')\operatorname{Re}\int_0^{t''} dt' G_b(t''-t')\hat{F}_b(t'). \tag{18}$$

To calculate the scattering spectrum $I(\omega)$, we use the quantum regression theorem [38, 39]

$$I(\omega) \sim \left|\mathbf{d}_{eg}\right|^2 \operatorname{Re} \int_0^{+\infty} \left\langle \hat{\sigma}^{\dagger}(t+\tau)\hat{\sigma}(t) \right\rangle e^{-i\omega\tau} d\tau, \qquad (19)$$

where the averaging is done over the noises $\hat{F}_b(t)$. As a result, we obtain

$$\begin{aligned}
I(\omega) = & \frac{2\omega_{vis}^4}{3c^3} \frac{\left|\mathbf{d}_{eg}\right|^2}{4} \left(\frac{\Omega_{vis}}{\omega_{vis}-\omega_0}\right)^2 \delta(\omega_{vis}-\omega) + \frac{2\omega_{IR}^4}{3c^3} \frac{\left|\mathbf{d}_{eg}\right|^2}{4} \left(\frac{\Omega_{IR}}{\omega_{IR}-\omega_0}\right)^2 \delta(\omega_{IR}-\omega) \\
& + \frac{\left|\mathbf{d}_{eg}\right|^2}{4} \left(\frac{\Omega_{vis}}{\omega_{vis}-\omega_0}\right)^2 \left[ \frac{2(\omega_{vis}-\omega_v)^4}{3c^3} \frac{g^2}{(\omega_0-\omega_{vis}+\omega_v)^2} \frac{(1+\bar{n}_v)\gamma_v}{(\omega_{vis}-\omega_v-\omega)^2+\gamma_v^2/4} \right. \\
& \left. + \frac{2(\omega_{vis}+\omega_v)^4}{3c^3} \frac{g^2}{(\omega_0-\omega_{vis}-\omega_v)^2} \frac{\bar{n}_v\gamma_v}{(\omega_{vis}+\omega_v-\omega)^2+\gamma_v^2/4} \right] \\
& + \frac{\left|\mathbf{d}_{eg}\right|^2}{4} \left(\frac{\Omega_{IR}}{\omega_{IR}-\omega_0}\right)^2 \left[ \frac{2(\omega_{IR}-\omega_v)^4}{3c^3} \frac{g^2}{(\omega_0-\omega_{IR}+\omega_v)^2} \frac{(1+\bar{n}_v)\gamma_v}{(\omega_{IR}-\omega_v-\omega)^2+\gamma_v^2/4} \right. \\
& \left. + \frac{2(\omega_{IR}+\omega_v)^4}{3c^3} \frac{g^2}{(\omega_0-\omega_{IR}-\omega_v)^2} \frac{\bar{n}_v\gamma_v}{(\omega_{IR}+\omega_v-\omega)^2+\gamma_v^2/4} \right].
\end{aligned} \qquad (20)$$

In the right-hand side of Eq. (20), the first two terms describe the Rayleigh scattering of both the coherent visible light beam of the probe with the frequency $\omega_{vis}$ and the coherent IR light of the pump with the frequency $\omega_{IR}$. One can see that both Stocks and anti-Stokes signals caused by the incidence of these signals are incoherent having the line halfwidth determined by the dissipation rate of nucleus vibrations, $\gamma_v$. The intensities of anti-Stokes signals are proportional to the number of quanta of thermal vibrations $\left\langle \hat{b}_0^{\dagger}(t)\hat{b}_0(t) \right\rangle = \bar{n}_v = \left( \exp(\hbar\omega_v/kT) - 1 \right)^{-1}$, while the intensities of the Stokes signals are proportional to $1+\bar{n}_v$. Since $\bar{n}_v \ll 1$, the ratio of intensities of these signals is $\exp(-\hbar\omega_v/kT)$.

Now, summing up incoherent, $\hat{b}_{incoh} = \hat{b}_0$, and coherent, $\hat{b}_{coh} = \hat{b}_2$, terms from Eq. (7), (11) and (16) we obtain (notice that $\hat{b}_1 = 0$):

$$\hat{b} = \hat{b}_{incoh} + \hat{b}_{coh} = \int_0^t G_b(t-t') \hat{F}_b(t') dt' + \frac{1}{4} \frac{g}{(\omega_v - 2\omega_{IR}) - i\gamma_v} \frac{\Omega_{IR}^2}{\omega_0^2 - \omega_{IR}^2} e^{-i2\omega_{IR}t}, \qquad (21)$$

where the first term, coming from Eq. (7) is the incoherent, thermal contribution $\hat{b}_{\text{incoh}}$ arising due to the noise $\hat{F}_b(t)$, the second term $\hat{b}_{\text{coh}}$, arising from Eq. (16), is the coherent contribution arising due to the external IR radiation.

In the case of the resonance, $\omega_{\text{IR}} = \omega_v/2$, taking into account that $\omega_{\text{IR}} \ll \omega_0$ we obtain the number of coherent quanta of the vibrational motion:

$$n_b^{\text{coh}} = |b_{\text{coh}}|^2 \approx \frac{1}{16}\left(\frac{\Omega_{\text{IR}}}{\omega_0}\right)^4 \left(\frac{g}{\gamma_v}\right)^2. \tag{22}$$

At the same time, the number of quanta of incoherent vibrations is equal to (see Ref. [38])

$$n_b^{\text{incoh}} = \left\langle \hat{b}_{\text{incoh}}^\dagger \hat{b}_{\text{incoh}} \right\rangle = \left\langle \int_0^t G_b^*(t-t')\hat{F}_b^\dagger(t')dt' \int_0^t G_b(t-t'')\hat{F}_b(t'')dt'' \right\rangle = \frac{1}{\exp(\hbar\omega_v/kT)-1}. \tag{23}$$

Using the typical values of the parameters for organic molecules, $\omega_v \approx 100$ meV, $\gamma_v \approx 1$ meV, $g \approx 10$ meV, $\omega_0 \approx 3$ eV, $\Omega_{\text{IR}} \approx 10$ meV, and $T = 20$ meV [50-52], we obtain $n_b^{\text{coh}} \sim 10^{-7} n_b^{\text{incoh}} \ll 1$. This means that the number of quanta of coherent vibrations is by many orders of magnitude smaller than the number quanta of the incoherent vibrational motion. Nonetheless, as we show below, coherent vibrations give the dominant contribution to the anti-Stokes signal.

In order to find the number of emitted photons, we should go further and consider the next order of the perturbation theory of Eqs. (2) and (3)

$$d\hat{\sigma}_3/dt + (i\omega_0 + \gamma_\perp)\hat{\sigma}_3 = -ig\hat{\sigma}_1(\hat{b}_2^\dagger + \hat{b}_2) + 2i\hat{\sigma}_1^\dagger \hat{\sigma}_1(\Omega_{\text{vis}}\cos(\omega_{\text{vis}}t) + \Omega_{\text{IR}}\cos(\omega_{\text{IR}}t)), \tag{24}$$

Equation (24) is linear with respect to $\hat{\sigma}_3$; therefore, the solution for $\hat{\sigma}_3$ can be found separately for each frequency. Since we are interested in the intensities of Stokes and anti-Stokes signals with frequencies $\omega_{\text{aSt}} = \omega_{\text{vis}} + 2\omega_{\text{IR}}$ and $\omega_{\text{St}} = \omega_{\text{vis}} - 2\omega_{\text{IR}}$, respectively, in the right-hand part of Eq. (24), we consider terms oscillating with frequencies $\omega_v \mp 2\omega_{\text{IR}}$ only:

$$d\hat{\sigma}_3/dt + (i\omega_0 + \gamma_\perp)\hat{\sigma}_3$$
$$= -\frac{i}{8}\frac{\Omega_{\text{vis}}\Omega_{\text{IR}}^2}{(\omega_{\text{vis}}-\omega_0)(\omega_0^2-\omega_{\text{IR}}^2)}\frac{g^2}{(\omega_v - 2\omega_{\text{IR}}) - i\gamma_v}e^{-i(\omega_{\text{vis}}+2\omega_{\text{IR}})t} - i\frac{\Omega_{\text{vis}}\Omega_{\text{IR}}^2}{\omega_0^2 - \omega_{\text{IR}}^2}e^{-i(\omega_{\text{vis}}+2\omega_{\text{IR}})t} \tag{25}$$
$$-\frac{i}{8}\frac{\Omega_{\text{vis}}\Omega_{\text{IR}}^2}{(\omega_{\text{vis}}-\omega_0)(\omega_0^2-\omega_{\text{IR}}^2)}\frac{g^2}{(\omega_v - 2\omega_{\text{IR}}) + i\gamma_v}e^{-i(\omega_{\text{vis}}-2\omega_{\text{IR}})t} - i\frac{\Omega_{\text{vis}}\Omega_{\text{IR}}^2}{\omega_0^2 - \omega_{\text{IR}}^2}e^{-i(\omega_{\text{vis}}-2\omega_{\text{IR}})t},$$

The solution of this equation is

$$\hat{\sigma}_3 = \frac{\Omega_{vis}\Omega_{IR}^2}{(\omega_0^2-\omega_{IR}^2)}\left[\frac{1}{8}\frac{g^2}{(\omega_{vis}-\omega_0)((\omega_v-2\omega_{IR})-i\gamma_v)}+1\right]\frac{\exp(-i(\omega_{vis}+2\omega_{IR})t)}{(\omega_{vis}+2\omega_{IR}-\omega_0)}$$
$$+\frac{\Omega_{vis}\Omega_{IR}^2}{(\omega_0^2-\omega_{IR}^2)}\left[\frac{1}{8}\frac{g^2}{(\omega_{vis}-\omega_0)((\omega_v-2\omega_{IR})+i\gamma_v)}+1\right]\frac{\exp(-i(\omega_{vis}-2\omega_{IR})t)}{(\omega_{vis}-2\omega_{IR}-\omega_0)},$$

(26)

From Eq. (26), the quantum regression theorem, Eq. (19), allows us to find the intensity of the coherent Raman signal:

$$I(\omega) = |\mathbf{d}_{eg}|^2 \frac{\Omega_{vis}^2\Omega_{IR}^4}{(\omega_0^2-\omega_{IR}^2)^2}\left|1+\frac{1}{8}\frac{g^2}{(\omega_{vis}-\omega_0)(\omega_v-2\omega_{IR}-i\gamma_v)}\right|^2$$
$$\times\left[\frac{2(\omega_{vis}+2\omega_{IR})^4}{3c^3}\frac{\delta(\omega_{vis}+2\omega_{IR}-\omega)}{(\omega_{vis}+2\omega_{IR}-\omega_0)^2}+\frac{2(\omega_{vis}-2\omega_{IR})^4}{3c^3}\frac{\delta(\omega_{vis}-2\omega_{IR}-\omega)}{(\omega_{vis}-2\omega_{IR}-\omega_0)^2}\right].$$

(27)

This intensity includes both Stokes and anti-Stokes coherent signals (the first and the second terms in the square brackets). If the resonance condition $\omega_{IR}=\omega_v/2$ is fulfilled, $I(\omega)$ takes the form:

$$I(\omega)=|\mathbf{d}_{eg}|^2\frac{\Omega_{vis}^2\Omega_{IR}^4}{(\omega_{vis}-\omega_0)^2\omega_0^4}\left|1-\frac{i}{8}\frac{g^2}{\gamma_v(\omega_0-\omega_{vis})}\right|^2\times$$
$$\times\left[\frac{2(\omega_{vis}+2\omega_{IR})^4}{3c^3}\delta(\omega_{vis}+2\omega_{IR}-\omega)+\frac{2(\omega_{vis}-2\omega_{IR})^4}{3c^3}\delta(\omega_{vis}-2\omega_{IR}-\omega)\right],$$

(28)

where the proportionality follows from Eq. (22).

As one can see from Eq. (28), for the scattering on coherent vibrational motion, the intensities of signals at the Stokes and anti-Stokes frequencies coincide. That distinguishes this result from the one for the scattering on the thermal vibrational motion, in which the ratio of these intensities is $\exp(-\hbar\omega_v/kT)$.

In Eq. (27), the term $g^2/8(\omega_{vis}-\omega_0)(\omega_v-2\omega_{IR}-i\gamma_v)$ has the resonance at $2\omega_{IR}=\omega_v$, while the rest of the expression has no resonances. The intensity of the resonant term is proportional to the number of coherent quanta of nucleus vibrations $n_b^{coh}$. From Eq. (26), one can see that the ratio of intensities of coherent Stokes and anti-Stokes peaks, defined by the coherent parts only, is about unity,

$$\frac{I_{St}}{I_{aSt}} = \left(\frac{\omega_{aSt} - \omega_0}{\omega_{St} - \omega_0}\right)^2 \approx 1, \qquad (29)$$

The reason for intensities of anti-Stokes and Stokes signals to be about the same is that coherent oscillations of nuclei cause classical modulation of the frequency of electronic system transition. This affects both signals in the same way. In incoherent Raman scattering, both anti-Stokes and Stokes signals arise due to the modulation of the electronic transition by thermal noises. In this process, the intensity of the incoherent Stokes signal is higher than that of the incoherent anti-Stokes signal.

Thus, in contrast to Eq. (20) in which the spontaneous Raman radiation is taking into account, Eq. (29) describes the coherent radiation, which is caused by stimulated radiation only. Consequently, the intensities of the Stokes and anti-Stokes signals are both proportional to $n_b^{coh}$ and are about the same. Thus, in contrast to CARS that allows for measuring the Stokes component only, in our approach, both signals can be measured.

The non-resonant term in (28) related to the electronic non-linearity. This non-resonant part of Eq. (28) corresponds to the non-resonant background also observed in CARS [53-57]. This background reduces the contrast of Raman signals. There are, however, some effective methods for the suppression of this background [58-62]. For example, it can be suppressed by the time delay between IR and visible light pulses [29, 63]. In view of this, special interest has the case of nitrogen, for which $\omega_v \approx 100$ meV, $\gamma_v \approx 0.05$ meV, $g \approx 50$ meV, $\omega_0 \approx 3$ eV and the ratio of resonant to non-resonant backgrounds are $\sim 10$. In such a case, it is not necessary to remove the non-resonant background.

According to the experimental data, the coherent Raman signal is proportional to the squared number of illuminated molecules, while the spontaneous Raman signal is linearly proportional to the number of molecules [64, 65]. Consequently, CARS can only be experimentally realized for extended samples. In this case, the phase relations between the incident and scattered waves become important. In a homogeneous medium, the wavevector mismatch between the incident fields and the induced polarization, $\Delta \mathbf{k} = \mathbf{k}_{aSt} - (\mathbf{k}_{vis} + 2\mathbf{k}_{IR})$, is caused by the temporal dispersion of the medium [36]. For non-zero $\Delta \mathbf{k}$, the radiating field will run out of phase with the induced polarization. Effective coherence of the vibrational motion of the molecules is maintained over the length scale of $L_c = 2\pi/\Delta \mathbf{k}$. In a gas medium, the temporal dispersion of the refractive index $n$ is small (for example, for nitrogen, $\Delta n \sim 10^{-5}$). Therefore in the case of the forward detected CARS ($\mathbf{k}_{aSt}$, $\mathbf{k}_{vis}$, and $\mathbf{k}_{IR}$ have the same directions), the coherence length for nitrogen is $L_c \sim 10^{-3}$ m. Further, the concentration of gas molecules, such as hydrogen or nitrogen is of the order of $\sim 10^{18} - 10^{19}$ cm$^{-3}$. Thus, in the coherence volume of about 1 mm$^3$, there are about $10^{16}$ molecules that are involved in the scattering process. For this reason, the coherent Raman signal dominates over the spontaneous signal. According to Eqs. (15) and (24), the intensity of the coherent Raman signal is proportional to $N^2 n_b^{coh}$, while the intensity of the spontaneous signal is proportional to $N n_b^{incoh}$. Hence, the intensity is enhanced by the factor $N n_b^{coh} / n_b^{incoh} \sim 10^8 - 10^9$. Considering that the spectrum of the coherent Raman scattering is narrower than that of incoherent scattering, and the signal of the former is more directional [29, 55], we have obtained a lower estimate for the ratio of intensities. Thus, one should expect that in our method, the coherent anti-Stokes signal is by four orders of magnitude greater than the incoherent anti-Stokes signal. In this sight, the special interest has a case of nitrogen molecules, for which at room temperature $T = 20$ meV, from Eqs. (26) and (27) one obtains $n_b^{coh} \sim 10 n_b^{incoh}$, and the coherent Raman signal exceeds the spontaneous one even for a single molecule.

To conclude, we note that the analytical results, which have been obtained within the perturbation theory, allow us to calculate the nonlinear dielectric permittivity $\chi^{(3)}$ that is usually used for the characterization of the Raman scattering. It can be obtained from the following arguments. The dipole moment of a single molecule is $d(t) = d_{eg} \langle \hat{\sigma}(t) \rangle$ [38, 39]. This gives that the specific polarization of the gas $P(t) = n d_{eg} \langle \hat{\sigma}(t) \rangle$ [39, 47], where $n$ is the concentration of the molecules. As it follows from Eq. (26), for the incident fields $E_{vis}$ having the frequency $\omega_{vis}$ and $E_{IR}$ with the frequency $\omega_{IR}$, the responses at the frequencies $\omega_{vis} \pm 2\omega_{IR}$ arise (see Supplemental Material for details):

$$P^{(3)}(\omega_{vis} \pm 2\omega_{IR}) = \frac{n d_{eg}^4}{\hbar^3 (\omega_0^2 - \omega_{IR}^2)(\omega_{vis} \pm 2\omega_{IR} - \omega_0)} \left[ \frac{1}{8} \frac{g^2}{(\omega_{vis} - \omega_0)((\omega_v - 2\omega_{IR}) \mp i\gamma_v)} + 1 \right] E_{vis} E_{IR}^2, \quad (30)$$

The corresponding the nonlinear susceptibilities $\chi^{(3)}(\omega_{vis} \pm 2\omega_{IR}; \omega_{vis}, \omega_{IR})$ are defined by the polarizations $P^{(3)}(\omega_{vis} \pm 2\omega_{IR})$, respectively. Thus,

$$\chi^{(3)}(\omega_{vis} \pm 2\omega_{IR}; \omega_{vis}, \omega_{IR}) = \frac{n d_{eg}^4}{\hbar^3 (\omega_0^2 - \omega_{IR}^2)(\omega_{vis} \pm 2\omega_{IR} - \omega_0)} \left[ \frac{1}{8} \frac{g^2}{(\omega_{vis} - \omega_0)((\omega_v - 2\omega_{IR}) \mp i\gamma_v)} + 1 \right], \quad (31)$$

The nonlinear susceptibilities $\chi^{(3)}(\omega_{vis} \pm 2\omega_{IR}; \omega_{vis}, \omega_{IR})$ have resonances at $2\omega_{IR} = \omega_v$.

In experiments, $\chi^{(3)}$ is not measured directly. It can be obtained the Raman cross section $\sigma_{Cross\_section}$ that can be calculated by integration of the spectrum of Stokes signal,

$$I_{St}(\omega) = \frac{2(\omega_{vis} - \omega_v)^4}{3c^3} \frac{|\mathbf{d}_{eg}|^2}{4} \left(\frac{\Omega_{vis}}{\omega_{vis} - \omega_0}\right)^2 \frac{g^2}{(\omega_0 - \omega_{vis} + \omega_v)^2} \frac{(1 + \bar{n}_v)\gamma_v}{(\omega_{vis} - \omega_v - \omega)^2 + \gamma_v^2/4}, \quad (32)$$

over all frequencies

$$\sigma_{Cross\_section} = \frac{1}{cE_{vis}^2/8\pi} \int_0^\infty I_{St}(\omega)d\omega. \quad (33)$$

This gives

$$\sigma_{Cross\_section} \approx \frac{4\pi}{3} k_{St}^4 \frac{|\mathbf{d}_{eg}|^4}{\hbar^2(\omega_{vis} - \omega_0)^2} \frac{g^2}{(\omega_0 - \omega_{St})^2}, \quad (34)$$

where $k_{St} = \omega_{St}/c$.

Finally, for $\chi^{(3)}$ we obtain.

$$\chi^{(3)}(\omega_{vis} - 2\omega_{IR}; \omega_{vis}, \omega_{IR}) = \frac{3}{32\pi} \frac{n}{\hbar k_{St}^4} \frac{(\omega_{vis} - \omega_0)(\omega_0 - \omega_{St})}{(\omega_0^2 - \omega_{IR}^2)((\omega_v - 2\omega_{IR}) + i\gamma_v)} \sigma_{Cross\_section}. \quad (35)$$

Note that similar relation between $\chi^{(3)}$ and $\sigma_{Cross\_section}$ exist for CARS as well [66].

**Discussion and conclusion**

Quanta of the vibrational motion are absorbed in Raman scattering via the anti-Stokes processes. The anti-Stokes signal, therefore, is proportional to the number of quanta of the nuclei vibrational motion. We propose a method of enhancement of the Raman signals for molecules with zero dipole moment transitions between vibrational states, based on an increase in the number of quanta of the vibrational motion due to illumination of molecules with IR coherent light.

Our method differs from the two-photon method suggested in Ref. [67], which is also requires an additional IR source. In the latter method it is also assumed that the dipole moment for the transitions between ground and excited vibrational states equals zero. Nevertheless, the excitation of the latter state is caused directly by of the incident IR radiation. This becomes possible thanks to a cascade of transitions between auxiliary vibrational states with *nonzero* dipole moments. This method is only applicable to specific molecules having such additional vibrational states.

The implementation of our method for Raman active molecules implies that a direct excitation of nuclei vibrations by incident IR light is not possible. The light nonresonantly affects the electronic subsystem, which via the nonlinear interaction excites nuclei vibrations. The enhancement of the Raman signal is caused by a parametric resonance interaction of the incident IR light with nucleus vibrations. It has also been shown that these processes can be considered by introducing nonlinear susceptibilities of the third order, $\chi^{(3)}$. The coherent Raman scattering can be considered as nonlinear four-wave mixing with $\chi^{(3)}(\omega_{vis} \pm 2\omega_{IR}; \omega_{vis}, \omega_{IR}, \omega_{IR})$. Since the coherent Stokes and anti-Stokes peaks are the results of nonlinear frequency mixing, their intensities are about the same and their ratio does not explicitly depend on temperature.

The coherent anti-Stockes signal is obtained in CARS. However, in contrast to our method, in CARS the stimulated Raman scattering can result in the energy transition between the laser beams [67]. This leads to signal suppression and limits the sensitivity [36]. In the method suggested here, the stimulated Raman scattering does not affect the sensitivity of the spectroscopy because neither Stokes nor anti-Stokes coincide with the frequencies of the incident fields.

The authors thank D. N. Kozlov and V. I. Fabelinsky for useful discussions and comments. A.A.L. acknowledges the support of the National Science Foundation under Grant No. DMR-1312707.